\documentclass{article}

\usepackage{microtype}
\usepackage{graphicx}
\usepackage{subfigure}
\usepackage{booktabs} 
\usepackage{amssymb}
\usepackage{amsfonts}
\usepackage{amsmath}[2000/07/18] 
\usepackage{amssymb}[2002/01/22] 
\usepackage{amsfonts}
\usepackage{bm}
\usepackage{xfrac}
\usepackage{stmaryrd}
\usepackage{mathtools}
\usepackage{eucal} 
\usepackage{multicol}
\DeclarePairedDelimiterX{\infdivx}[2]{(}{)}{%
	#1\;\delimsize\|\;#2%
}

\DeclareMathOperator*{\argmax}{arg\,max}  
\usepackage{hyperref}

\newcommand{\name}{Cutoff SGD }

\usepackage[accepted]{icml2018}

\icmltitlerunning{High Throughput Synchronous Distributed Stochastic Gradient Descent}

\begin{document}

\twocolumn[
\icmltitle{High Throughput Synchronous Distributed Stochastic Gradient Descent}



\icmlsetsymbol{equal}{*}

\begin{icmlauthorlist}
\icmlauthor{Michael Teng}{ox}
\icmlauthor{Frank Wood}{ox}
\end{icmlauthorlist}

\icmlaffiliation{ox}{Department of Engineering Sciences, University of Oxford, Oxford, United Kingdom}

\icmlcorrespondingauthor{Michael Teng}{mteng@robots.ox.ac.uk}

\icmlkeywords{Machine Learning, ICML}

\vskip 0.3in
]



\printAffiliationsAndNotice{} 

\begin{abstract}
We introduce a new, high-throughput, synchronous, distributed, data-parallel, stochastic-gradient-descent learning algorithm.  This algorithm uses amortized inference in a compute-cluster-specific, deep, generative, dynamical model to perform joint posterior predictive inference of the mini-batch gradient computation times of all worker-nodes in a parallel computing cluster.  We show that a synchronous parameter server can, by utilizing such a model, choose an optimal cutoff time beyond which mini-batch gradient messages from slow workers are ignored that maximizes overall mini-batch gradient computations per second.  In keeping with earlier findings we observe that, under realistic conditions, eagerly discarding the mini-batch gradient computations of stragglers not only increases throughput but actually increases the overall rate of convergence as a function of wall-clock time by virtue of eliminating idleness.  The principal novel contribution and finding of this work goes beyond this by demonstrating that using the predicted run-times from a generative model of cluster worker performance to dynamically adjust the cutoff improves substantially over the static-cutoff prior art, leading to, among other things, significantly reduced deep neural net training times on large computer clusters.  

%
\end{abstract}

\section{Introduction}
\label{submission}

Deep learning success stories are predicated on large neural network models being trained using ever larger amounts of data.  While the computational speed and memory available on individual computers and GPUs grows ever larger, there always will remain some problems and settings in which the amount of training data available will not fit entirely into the memory of one computer.  What is more, and even for a fixed amount of data, as the number of parameters in a neural network or the complexity of the computation it performs increases, so too does the time it takes to train.  Both large training data and complex networks inspire parallel training algorithms.

In this work we focus on parallel stochastic gradient descent (SGD).  Like the substantial and growing body of work on this topic (e.g.~non-exhaustively: \citet{recht2011hogwild,dean2012large,mcmahan2014delay,zhang2015deep}) we too will focus on gradient computations computed in parallel on ``mini-batches'' drawn from the training data.  However, unlike most of these methods which are {\em asynchronous} in nature, we focus instead on improving the performance of {\em synchronous} distributed SGD, very much like \citet{chen2016revisiting}, upon whose work we directly build.  

The main problem in fully synchronous distributed SGD is the \textit{straggler effect}.  This real-world effect is caused by the small and constantly varying subset of worker nodes that, for whatever random reasons, always perform their mini-batch gradient computation slower than the rest of the concurrent workers, causing long idle times in all of the workers which have already finished. \citet{chen2016revisiting} introduced a method of mitigating the straggler effect on wall-clock convergence rate by picking a fixed cut-off for the number of workers on which to wait  before synchronously updating the parameter on a centralized parameter server.  They found, as we demonstrate in this work as well, that the increased gradient computation throughput that comes from reducing idle time more than offsets the loss of a small fraction of mini-batch gradient contributions per gradient descent step.


Our work exploits this same key idea but substantially improves the way the likely number of stragglers is identified.  In particular we instrument and generate training data once for a particular compute cluster then use it to train a lagged generative latent-variable time-series model that encodes the joint worker run-time behavior of all the workers in the cluster.  For highly contentious clusters with poor job schedulers, such a model might reasonably be expected to learn to model latent-states that produce correlated, grouped increases in observed run-times due to resource contention.  For well-engineered clusters such a model might learn that worker run-times are nearly perfectly independently and identically distributed.  

Specifying such a flexible model by hand would be difficult.  Also, as we will soon explain, we will need to perform real-time posterior predictive inference in said model at distributed synchronous SGD run-time to dynamically predict straggler cut-off.   For both these reasons we use the variational autoencoder loss \citep{kingma2013auto} to simultaneously learn not only the model \citep{krishnan2017structured} parameters but also the parameters of an amortized inference neural network \citep{ritchie2016deep,le2016inference} that allows for real-time approximate predictive inference of worker run-times.



The main contributions of this paper are:
\begin{itemize}
	\item The idea of using amortized inference in a deep latent-variable time-series model to predict computer cluster worker run-times, in particular for use in a distributed synchronous gradient descent algorithm.
	\item The dynamic cut-off distributed synchronous gradient descent algorithm itself, including in particular the approximations made to enable real-time posterior predictive inference.
	\item The empirical verification at scale of the decrease in time to convergence that our algorithm yields when training deep neural networks.
\end{itemize}
 
 The rest of the paper is organized as follows.  Section 2. grounds our investigation in its motivation of synchronous SGD speedup. Section 3. outlines two models of compute cluster run-times used to dynamically determine a straggler cut-off. Section 4. highlights our experimental results.

\section{Background and Motivation}

In stochastic gradient descent, we rely on unbiased estimates of the gradient in order to update the global parameter settings. 
Distributed mini-batch SGD differs from serial mini-batch SGD in that the mini-batch of size $m$ is distributed to $n$ worker computers that locally compute sub-mini-batch gradients before communicating the result back to a centralized parameter server that updates the parameter using a gradient update step:
\begin{equation}
\label{eqn:sgd}
\theta^{(t+1)} = \theta^{(t)} - \alpha \frac{1}{n} \sum_{i=1}^n  [f(\theta^{(t)}, (i - 1)\frac{m}{n}, i\frac{m}{n} ] 
\end{equation}
with 
\[f(\theta, a, b) = \frac{1}{b - a} \sum_{z=0}^{b - a} \nabla_{\theta^{(t)}} F(\theta, x^{(z)}, y^{(z)}) \]
where $F$ is the loss function and $\alpha$ is the learning rate.  Distributed SGD, as shown, uses unbiased gradient estimates, leaving $\alpha$ and $m$ as tunable hyperparameters governing the convergence properties of the algorithm, similar to the single-threaded case. Too high a learning rate causes the algorithm to diverge, while low mini-batch size and low $\alpha$ both can produce convergence to local minima \cite{hoffer2017train}.

\begin{figure}[tbpb]
	\begin{center}
		\includegraphics[width=65mm]{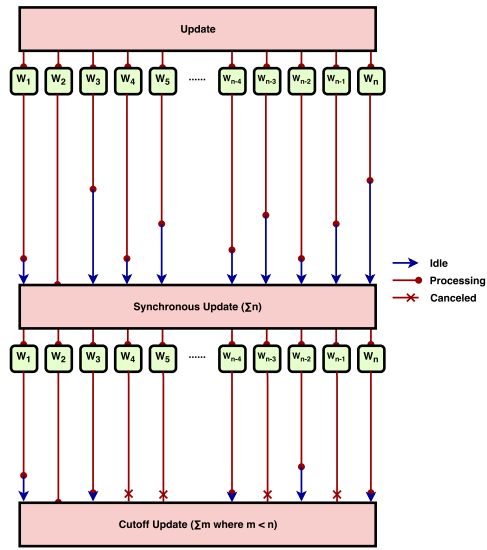}
		\caption{Diagram highlighting source of low throughput arising from stragglers.  From top to bottom: synchronous gradient updates require waiting for all workers to finish.  Alternatively, we propose to predict the number of stragglers that leads to highest overall throughput in terms of number of centralized gradient updates per unit time.  The mechanism we employ is to use a model to predict the joint ordered run-times of all workers, eagerly dropping a fraction of straggling workers.}
		\label{icml-historical}
	\end{center}
\end{figure}

\subsection{Effect of Stragglers }

In synchronous SGD, we can attribute low throughput in the sense of central parameter updates per unit time to the straggler effect that arises in real-world cluster computing scenarios with multiple workers computing in parallel. Consider Equation~\ref{eqn:sgd}, in which $f(\theta)$ is computed independently on an memory-isolated logical processor.  Let $x_j$ be the time it takes for $f$ to be computed on the worker indexed by $j$ for $j \in {1...n}$.  Distributed computers are not ideal, otherwise $x_j$ would be a constant, independent of $j$, and all workers would finish at the same time and only be idle while the parameter server aggregates the gradients and sends back the new parameters.  Instead $x_j$ is actually random.  Moreover, the joint distribution of all the $x_j$'s is likely, again in real-world settings, to be non-trivially correlated owing to cluster architecture, etc.  For instance most modern clusters consist of computers or graphics processing units each in turn having a small number of independent processors, so slow downs in one logical processing unit are likely to be exhibited by others sharing the same, for instance, bus or network address.  What is more, in modern operating systems, time-correlated contention is quite common, particularly in clusters under queue management systems, when, for instance, other processes, operating system or user, are concurrently executed.  All this yields worker compute times that may be non-trivially correlated in both time and in ``space.''  

Our aim is to significantly reduce the effect of stragglers on throughput and to do so by modeling cluster worker compute times in a way that intelligently and adaptively responds to the kinds of correlated run-time variations actually observed in the real world.  What we find is that doing so improves overall performance of distributed mini-batch SGD. 
\begin{algorithm}[tb]
	\caption{\name}
	\label{alg:cutoff}
	\begin{algorithmic}[1]	
		\STATE {\bfseries Input:} data $ X$, \\ $~~~~~~~~~~~$ size $m$, \\$~~~~~~~~~~~$ inference network $q_\phi$, \\$~~~~~~~~~~~$ learning rate $\alpha$

		\IF{worker, $w$}
		
		\FUNCTION {compute\_gradient($\theta$)}
		\STATE $s_w \gets$ timestamp()
		\FOR{$i=1$ {\bfseries to} $b$}
		\STATE $x^{(z)}, y^{(z)} \gets$ sample($ X$)
		\STATE $f_{w} \gets f_{w} + \frac{1}{m}  \nabla F(\theta, x^{(z)}, y^{(z)})$
		\ENDFOR
		\STATE $r_w \gets$ timestamp() $- s_w$
		\STATE \textbf{send} $f_{w}$, $r_w$ $\rightarrow$  parameter server
		\ENDFUNCTION
		\WHILE{\textbf{not} converged}
		\STATE {\bfseries spawn} thread {\bfseries run} compute\_gradient
		\STATE {\bfseries await} \textbf{receive} $\theta' \gets$ parameter server	
		\STATE $\theta \gets \theta'$
		\STATE \textbf{terminate} thread
		\ENDWHILE
		\ENDIF		
		\IF{parameter server}
		\STATE obs $\gets $ \{\}
		\WHILE{\textbf{not} converged}
		\STATE $G \gets $ \{\}, 
		\STATE $c \gets$ predict\_cutoff($q_\phi$, obs)
		\WHILE{$j < c$}
		\STATE \textbf{receive} $f_w$, $r_w$ $\gets$ any worker
		\STATE obs $\gets$ obs $\cup ~{r_w}$
		\STATE $G \gets G \cup {f_w}$
		\ENDWHILE
		\STATE $\theta' \gets \theta - \alpha \frac{1}{c} \sum_{f\in G} f$
		\STATE \textbf{send} $\theta' \rightarrow$ all workers
		\ENDWHILE
		\ENDIF
		
	\end{algorithmic}
\end{algorithm}

\section{Methodology}

Our approach works by maximizing the total throughput of parameter updates during a distributed mini-batch SGD run. The basic idea, shared with \citep{chen2016revisiting}, is to predict a cutoff, $c < n$, (Alg.~\ref{alg:cutoff}, line 23) for each iteration of SGD which dictates the total number of workers on which to wait before taking a gradient step in the parameter space (Alg.~\ref{alg:cutoff}, line 29). To be concrete about why we want to do this: if the slowest straggler takes 10 seconds to finish, but the second slowest takes 8, then there is already a 20\% reduction in the wall-clock time simply by setting the cutoff to $n - 1$.

The central considerations are: what is the notion of throughput we should optimize?  And how do we predict the cutoff that achieves it?

Simply optimizing overall run-time admits a trivial and unhelpful solution of setting the cutoff to be all workers. Each iteration and the overall algorithm would then take no time.  Instead we seek  to maximize the number of workers to finish in a given amount of time, i.e.~throughput $\Omega(c)$, which we define to be:
\[\Omega(c) = \frac{c}{\tilde x_{(c)}}\]
where $c$ indexes the {\em ordered} worker run-times $\tilde x_{(c)}$.  Note that, for now and throughout when clear, we will avoid indexing run-times by SGD loop iteration, although we specifically will make use of temporal correlation between worker run-times soon enough.

We define our objective to be maximizing the throughput of the system as defined above, i.e.~$\argmax_{c} \Omega(c)$, which, as it turns out, will yield improved overall learning as a consequence of calculating and incorporating the maximum number of gradients over time. 
%
%
%

Setting the cutoff optimally and dynamically requires a model which is able to learn and  predict the joint run-times of all cluster workers.   With such a model, we aim to make highly informed and accurate predictions about the next set of run-times per worker and consequently make a real-time optimal choice of $c$ for the subsequent loop of sub-mini-batch gradient calculations.  How we model computer cluster worker performance follows.

\subsection{Modeling Computer Cluster Worker Performance}	

As before, let $x_j\in\mathbb{R}^+$ be the random times it takes for $f$ to be computed on the worker indexed by $j$.  Assume that these are distributed according to some distribution $p$.

\subsubsection{Order Statistics}

Given a set of $n$ identically $p$-distributed random variables $x_1, x_2,\ldots , x_n$ we wish to know the joint distribution of the $n$ {\em sorted} random variables $\tilde{x}_{(1)}, \tilde{x}_{(2)}, \ldots, \tilde{x}_{(n)}$.  Such quantities are known as ``order statistics.''  For instance under the assumption that $x_j = \mathcal{N} (\mu_{x},\sigma^2_{x})$ the distribution of each order statistic $p(\tilde{x}_{(1)}), p(\tilde{x}_{(2)}), ... , p(\tilde{x}_{(n)})$ is independent and $\mathbb{E}[\tilde{x}_{(1)}] \leq \mathbb{E}[\tilde{x}_{(2)}], ... , \leq \mathbb{E}[\tilde{x}_{(n)}]$. Each $p(\tilde{x}_{(j)})$ describes the distribution of the $j^{th}$ largest sorted run-time under independent draws from this underlying distribution. 

Under the given independent and identically distributed (iid) normality assumption the distribution of the each order statistic has closed form:
\[
 p(\tilde{x}_{(j)}) = Z(n,j)\int_{-\infty}^{\infty}{x}[\Phi({x})]^{j-1}[1 - \Phi({x})]^{n-j} p({x})d{x}
\]
where $\Phi({x})$ is the cumulative distribution function (CDF) of $\mathcal{N} (\mu_{t},\sigma^2_{ t})$ and $Z(n,j) = \frac{n!}{(j - 1)! (n-j)!}$
Note that each order statistic's distribution, including the maximum, increases as the variance of the run-time distribution increases, while the average run-time does not. 

 \begin{figure*}[tbh]
 	\begin{center}
 		\centerline{\includegraphics[width=.70\textwidth]{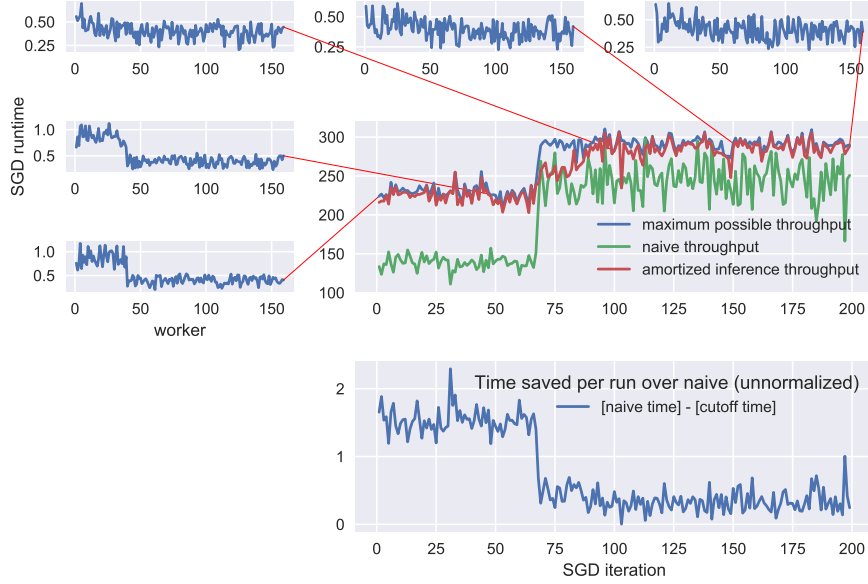}}

 		\caption{Results of throughputs given by amortized inference. Each runtime plot (5 surrounding the top figure) shows the individual runtimes of the worker indexed by the x-axis during an iteration of SGD on our local cluster. We highlight SGD iterations 1, 50, 100, 150, and 200 which highlight two significantly different regimes of persistent time and machine identity correlated worker runtimes. The top large figure displays a comparison of throughputs achieved by waiting for all workers to finish (green) and using the inferred cutoff method (red) relative to the ground truth maximum achievable (oracle). The bottom figure displays the reduction in time per iteration when \name is used.}
 		\label{fig:throughputs}

 	\end{center}
 \end{figure*}

Given $n$ workers, the expected average idle time for each worker if synchronizing on all completing can be derived to be:

\begin{equation}
\label{3}
\frac{1}{n}\sum_{j = 1}^{n} \mathbb{E}[\tilde{x}_{(n)} - \tilde{x}_{(j)}] \approx \mathbb{E}[{\tilde{x}_{(n)}}] - \mathbb{E}[{\tilde{x}_{(\frac{n}{2})}}]
\end{equation}

The latter approximation holds because the order statistics of iid draws from a Gaussian are both independent and symmetric around the middle order statistics; workers wait on average the difference of the longest (highest) order statistic and the middle order statistic. 

As a baseline in subsequent sections we will use a useful approximation of the expectations of order statistics under this iid normality assumption.  This is known as the Elfving (1947) formula \cite{royston1982algorithm}:
	
\begin{equation}
\mathbb{E}[\tilde{x}_{(j)}] \approx \mu_{t} + \Phi^{-1}\left(\frac{n-\frac{\pi}{8}}{j-\frac{\pi}{4}+1}; 0, 1\right)\sigma_{t} \label{eqn:elfving}
\end{equation}

It is not known how to derive the analytic form of the joint order statistic distribution of non-Gaussian distributed correlated random variables.  However a Monte Carlo approximation of the order statistics is straightforward: use a model to predict the joint distribution of the $x_j$'s, then sample, sort, and record the values of all $n$ sorted samples, and then repeat.  Towards that end we will first develop a model of correlated compute times from which we will then be able to construct Monte Carlo order statistic estimates for use in determining the optimal cutoff threshold.

\subsubsection{Generative Model}

Before introducing the design of the generative model we use to predict worker run-times, first consider why a generative model here is nearly absolutely necessary, certainly in comparison to a purely autoregressive model, for predicting run-times given a lagged window.  In short we can only consider  worker run-time prediction models that are extremely sample efficient to train.  This is because one receives no benefit whatsoever if the predictive model should require collected training data for many thousands of distributed SGD runs before being able to use it.   We also can only consider a kind of model that allows real-time prediction because it will be in the inner loop of the parameter server and used predict at run-time how many straggling workers to ignore.  Deep neural net auto-regressors satisfy the latter but not the former.  Generative models satisfy the former but historically not the latter; except now deep neural net guided amortized inference in generative models does.  This forms the core of our technical approach. 


We will model the time sequence of observed joint worker run-times $\bm x_{T-\ell}, \ldots, \bm x_T$ using a hidden Markov model where $\bm z_{T-\ell}, \ldots, \bm z_T$ is the time evolving unobserved latent state of the cluster.  The dependency structure of our model factorizes as:

\[
p_\theta(\bm x_{{T-\ell}:T}, \bm z_{{T-\ell}:T}) = \prod_{i={T-\ell}}^{T}p_\theta(\bm z_i | \bm z_{i - 1})\prod_{i = {T-\ell}}^{T} p_\theta(\bm x_i | \bm z_i)
\]

where, for reasons specific to amortizing inference, we will restrict our model to a fixed-lag $\ell$ window. 
The principal model use is the accurate prediction of the next set of worker run-times from those that have come before:
%
{\small
\begin{align}
&p(\bm x_{T+1}| \bm x_{{T-\ell}:T}) = \label{eqn:postpredtrue}\\
&\int p_\theta( \bm x_{T + 1} | \bm z_{T + 1})p_\theta(\bm z_{T+ 1} | \bm z_{T}) p(\bm z_{{T-\ell}:T} | \bm x_{{T-\ell}:T}) d\bm z_{{T-\ell}:T+1} \nonumber
\end{align}
}

\subsubsection{Model Learning and Amortized Inference}

With the course-grained model dependency defined, it remains to specify the fine-grained parameterization of the generative model, to explain how to train the model, and to show how to perform real-time approximate inference in the model.  

First we  use the deep linear dynamical model introduced by \cite{krishnan2017structured}. Namely, the transition and emission functions in our model are parametrized by neural networks:

\[\bm z_{t} \sim \mathcal{N}(G_{\theta}(\bm z_{t - 1}), H_{\theta}(\bm z_{t - 1})))\]
\[\bm x_{t} \sim \mathcal{N}(I_{\theta}(\bm z_{t}), J_{\theta}(\bm z_{t})))\]

 whose specific architecture is:

\[I_{\theta}(\bm z_{t}) = MLP_2(\bm z_{t-1},Identity, Identity) \] 
\[J_{\theta}(\bm z_{t}) = MLP_2(I_{\theta}(\bm z_{t}), ReLU, Softplus) \]

where $I$ is the identity matrix and $MLP_n(l, A, B, ...)$ denotes an $n$-layer multilayer perceptron containing nonlinearities, $A$, $B$, etc.

Our model also utilizes the gated transition function for $G$ and $H$:
\[G_{\theta}(\bm z_{t - 1}) = (1 - g_t) \cdot MLP_1(\bm z_{t- 1},Identity) + g_t \cdot h_t\] 
\[H_{\theta}(\bm z_{t - 1}) = MLP_1(ReLU(G_{\theta}(\bm z_{t - 1})), Softplus)\]
\[g_{t} = MLP_2(\bm z_{t-1},ReLU, Sigmoid) \]
\[h_{t} = MLP_2(\bm z_{t-1},ReLU, Identity). \]

  The flexibility of such a model allows us to avoid making restrictive or inappropriate assumptions that might be quite far from the true generative model while imposing rough structural assumptions that seem appropriate like correlation over time and correlation between workers at a given time.

The remaining tasks are to, given a set of training data, learn $\theta$ {\em and} train an amortized inference network to perform realtime inference in said model.  For this we utilized the variational autoencoder-style loss used for amortized inference in deep probabilistic programming with guide programs \citep{ritchie2016deep}.  The guide program structure we used is a structured left-right model:
\[q_{\phi}(\bm z_t | \bm z_{T-\ell:t}, \bm x_{T-\ell:T}) = \prod_{j=1}^{N}\mathcal{N}(\mu_{q_\phi}(j), \sigma_{q_\phi}(j)) \]

where $*$ denotes scalar multiplication below:
\begin{align*}
\mu_{q_\phi} &= K_\phi(h_{out}) = MLP_1(h_{out},Identity) \\
\sigma_{q_\phi} &= L_\phi(\mu_{q_\phi}) = MLP_1(\mu_{q_\phi},Softplus) \\
 h_{out} &= \frac{1}{3} * (MLP_1(\bm z_{t - 1}, Tanh) + h_{left} + h_{right})\\
 h_{left} &= RNN(\bm x_{T-\ell:t - 1}, ReLU)\\
 h_{right} &= RNN(\bm x_{t+1:T}, ReLU)
\end{align*}

We use stochastic gradient descent to simultaneously optimize the variational evidence lower bound (ELBO) with respect to both $\phi$ and $\theta$: 
{
\begin{align}
\textrm{ELBO} &=  \mathbb{E}_{q_\phi(\bm z_{T-\ell:t} | \bm x_{T-\ell:T})}\log\left(\frac{p_\theta(\bm x_{T-\ell:t}, \bm z_{T-\ell:t})}{q_\phi(\bm z_{T-\ell:t} | \bm x_{T-\ell:T})}\right) \nonumber 
\end{align}
}
where \[q_\phi(\bm z_{T-\ell:t} | \bm x_{T-\ell:T}) = \prod_{t={T-\ell}}^T q_{\phi}(\bm z_t | \bm z_{T-\ell:t}, \bm x_{T-\ell:T}).\]
Doing this yields an extremely useful by-product.  Maximizing the ELBO also drives the KL divergence between $q_\phi(\bm z_{T-\ell:t} | \bm x_{T-\ell:T})$ and $p_\theta(\bm z_{T-\ell:t} | \bm x_{T-\ell:t})$ to be small.  We will exploit this fact in our experiments to speed run-time prediction.

In particular we will directly approximate Equation~\ref{eqn:postpredtrue} by
{\small
\begin{align}
&p(\bm x_{T+1}| \bm x_{{T-\ell}:T}) \nonumber \approx\\
&\int p_\theta( \bm x_{T + 1} | \bm z_{T + 1})p_\theta(\bm z_{T+ 1} | \bm z_{T}) q_\phi(\bm z_{{T-\ell}:T} | \bm x_{{T-\ell}:T}) d\bm z_{{T-\ell}:T+1} \nonumber\\
&\approx \frac{1}{K}\sum_{k=1}^{K} p_\theta( \bm x_{T + 1} | \bm z_{T + 1})p_\theta(\bm z_{T+ 1} | \bm z_{T}^{(k)})
\end{align}
}

with $\bm z_{T}^{(k)}$ being the last-time-step marginal of the $k$th of $K$ samples from $q_\phi(\bm z_{{T-\ell}:T} | \bm x_{{T-\ell}:T})$.

%
%
%
%
%

During training, and at test-time, we normalize the observations by dividing out the 2 times the mean of the first fixed-lag window. In doing so, we avoid retraining the model for neural networks and batch sizes that cause longer runtimes.

  \begin{figure*}[tb]
  	\vskip 0.2in
  	\begin{center}
  		\centerline{\includegraphics[width=.80\textwidth]{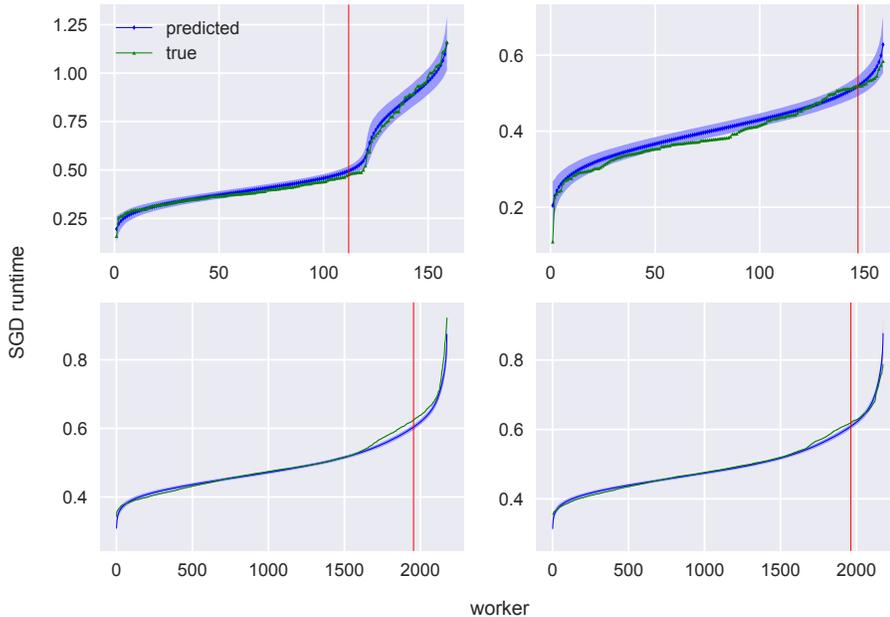}}

  		\caption{ Runtime profiles of various iterations of SGD of the validation set in our training step. The maximum throughput cutoff under the model predictions is shown in red, indicating a large chunk of idle time is reduced as a result of stopping early. Top: selected observed runtimes vs predicted runtime order statistics for 158 worker count model. Notably, when there are exceptionally slow workers present, the cutoff is set to proceed without any of them as seen in the top-left figure. Bottom: selected runtime comparisons for 2175 worker count model. All predicted order statistics are shown with $\pm 2$ standard deviations}
  		\label{fig:worker-runtimes}

  	\end{center}
  	\vskip -0.2in
  \end{figure*}
  


%
%
%
%

\section{Experiments}

In all experiments we use a twenty-timestep lag, i.e.~$\ell = 20$

\subsection{Predicting Worker Runtimes}

To test our model's ability to accurately predict joint worker runtimes sufficiently well that the approximate order statistics derived by {\em sorting} the output of the proposal network $q_\phi$ match the true ordered runtimes of workers in subsequent timesteps, we perform the same experiment on two clusters of distinctly different architectures and sizes.  
In particular we record worker compute times for fully synchronous SGD iterations while training a deep neural network.   Using these we  train our generative model parameters, the proposal network, then use both to make predictions about next time-step worker processing times and compute the cut-off we would use at run-time and compare the predicted runtimes and cut-offs to the ground-truth observed and computed from the known actual next-time-step worker runtimes.

On one cluster comprised of four nodes of forty logical Intel Xeon processors, we trained a 3-Layer CNN network to do MNIST classification with a 60000:10000 training to validation data split. We use a single parameter server leaving an available 158 worker count, across which we trained using synchronous SGD and recorded each worker runtime of an iteration of SGD for 1 hour. The resulting mean and standard deviation of the worker runtimes were 1.057 and 0.393 seconds respectively. 




Using these values in the Elfving formula (Eqn.~\ref{eqn:elfving}), we find that the maximum expected runtime out of 158 independent workers would be 2.1063. This means that approximately on average, in fully synchronous SGD, workers are spending 1.049 seconds idle while the longest running thread finishes its computation. During this time, a second gradient could almost be calculated, which provides some insight into the large increase in efficiency of our approach. 

In Figure~\ref{fig:throughputs} there is clear evidence that the strong assumptions of independence and identically distributed runtimes required to use the iid normality assumption do not hold. Figure~\ref{fig:throughputs} clearly shows what can happen on a highly contentious cluster, which produced different levels of correlated worker runtimes throughout the hour. In order to reduce the total wait times, a model of a compute cluster is required, and in particular one that does not make unnecessary and inaccurate assumptions about the distributions from which the runtimes are distributed.

The crux of our approach is to reduce the wait time of these processes.  Thus we trained our model and inference network using Adam with gradient clipping on the data collected from the small cluster. In addition, we also trained a production scale inference network with data taken from a Cray XC40 supercomputer operating on 32 KNL nodes with 68 logical cores per node. The worker counts available to us on these systems are 158 and 2175, respectively.

Both trained models display high performance on validation sets, where we test by comparing the next available vector of run-times against the predicted run-times emitted by the preceding 20 timestep sequence (see Figure~\ref{fig:worker-runtimes}). On our local cluster, we also discovered a set of slow workers that persisted for about 1000 iterations likely due to the contention of resources by another unrelated job that overlapped our training data collection phase. In the first mode, lasting from iterations 1 to 61, we observe a single slow compute node (the cluster in question contains 4 nodes with 40 cores each) as the main bottleneck. In the second mode, this slow machine equilibrates with the remaining 3, and our run-times are more uniform throughout the workers. We include this window in our test data and demonstrate that the model has learned both dynamical modes.

In Figure~\ref{fig:throughputs}, we compare the throughputs achieved by our approach with the maximum and naive throughputs in our 158 worker model. Again, our validation set was chosen to include a window of interest where at iteration 61, the compute cluster sheds a set of 40 slow nodes and operates at uniform efficiency. During this transition, our inferred cutoff is only set suboptimally for a few iterations, before recovering near maximum performance after iteration 85.

\subsection{Handling Censored Run-times}

As described, we use the learned inference network to predict future cutoffs rather than the generative model. Because variational inference jointly learns the model parameters along with the inference network, we could theoretically use an inference algorithm such as SMC \cite{doucet2001introduction} for more accurate estimates of the true posterior. However, our cutoff prediction must be done in an amortized setting, because we rely on it to be set for a gradient run prior to the updates returning from the workers. In a setting which requiring fast, repeated inference, using an amortized method is often the only approach, especially in large complex models. 

However, when using amortized inference, there is a practical implication of dealing with partially observed and in fact censored data. Since at run-time we are only waiting for $c$ gradients up to the cutoff, and are in fact actually killing the straggling workers, we do not have the run-time information from the straggling workers that would have finished past the cutoff. This results in censored observations, and we know that censoring occurs right at $\tilde f_{(c)}$. Inference in the generative model could directly be made able to deal with censored data, however our inference network runs an RNN which was trained on fully observed run-time vectors and therefore requires fully observed input to function correctly. Because of this, we describe an effective approximate technique for imputing the missing worker runtime values. 

Our practical solution is to sample a new uncensored data point for every worker whose gradients are dropped. Because we push estimates of the approximate posterior through the generative model, we have a predictive run-time distribution for the \textit{current} iteration of SGD before receiving actual updates from any worker. When eventually the cutoff is reached, and the corresponding rate censor is observed, we are left with run-time distributions, which are left truncated at $\tilde x_{(c)}$:

\[
p(\tilde x; \tilde x > \tilde x_c) = \frac{p(\tilde x)}{\int_{\tilde x_{(c)}}^{\infty} p(\tilde x) d{\tilde x} }
\]
where we have left off the time index for clarity and $\tilde x$ is any one of the censored worker runtime observations. 

When a censored value is required, we take its corresponding predicted run-time distribution and sample from the right tail truncated distribution to get an approximate value for that missing run-time. We find that this method works well to propagate the model forward, leading to still accurate predictions.

\subsection{ Wall Clock Speedup}

\begin{figure}[t]
	\begin{center}
		\centerline{\includegraphics[width=\columnwidth]{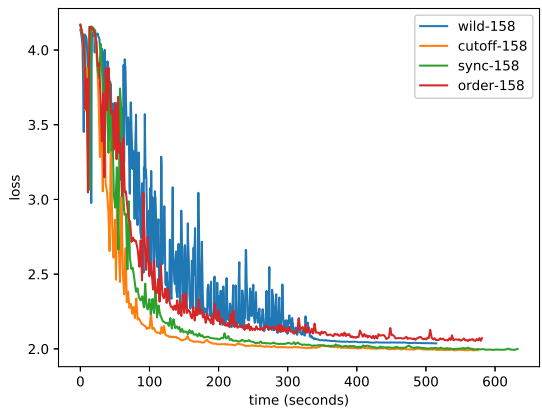}}
		\caption{MNIST validation loss convergence for 4 methods: wild=Hogwild, cutoff=our algorithm, sync=fully synchronous distributed SGD, order=cutoff based on analytic normality cutoff criterion.  Batch size - 10112, learning rate scaled to 0.64 for sync and 0.004 (0.64 / num workers) for async.  Cutoff achieves lower loss faster than all other methods, and finishes faster than all other synchronous-family methods too.}
		\label{fig:convergence-rate-comparison}
	\end{center}
\end{figure}
We report results for the simple MNIST example run on the 160 node computer cluster. All distributed cutoff SGD experiments were run with sampling a mini-batch with replacement. For some other distributed SGD implementations, a subset of the data is pre-partitioned onto each worker to save networking cost. However, here we cannot do that because if some workers remain inherently slow  then their gradients will always be dropped as a result of maximizing throughput.   All experiments use a single parameter server, which did not present a bottleneck during testing.  We implement the popular asynchronous SGD algorithm, Hogwild, in order to compare the convergence of the noise-adding, but perhaps faster wall-clock training rate, of an asynchronous method. Hogwild's algorithm has the parameter server communicating with the workers at each update, while synchronous SGD allows for only small communication bandwidth to report a rate once finished. When the parameter server is able to set the cutoff, it broadcasts this list of participants to its workers as a bit array, and then workers who do not finish zero their gradients and the full array performs its update locally after sharing information in an all ring reduce. This method also lowers the communication requirements, an optimization that is unachievable in the most asynchronous implementations. 

Figure~\ref{fig:convergence-rate-comparison} shows that our method achieves the fastest convergence to the lowest lost among comparison methods performing synchronous SGD. Hogwild outperforms our approach in wall-clock time, but its convergence is to a higher validation loss.

\section{Discussion}

We have presented an improved, faster way to do synchronous distributed gradient descent. Our primary contributions include describing how a model of worker runtimes can be used to predict order statistics that allow for a near optimal choice of straggler cutoff that maximizes gradient computation throughput. 

While the focus throughout has been on on vanilla SGD, it should be clear that our method and algorithm can be nearly trivially extended to most optimizers of choice so long as they are stochastic in their operation on the training set. Most methods for learning deep neural network models today fit this description, including for instance the Adam optimizer \cite{kingma2014adam}.

We conclude with a note that our method implicitly assumes that every minibatch is of the same computational cost in expectation, which may not always be the case. Future work could be to extend the inference network further \cite{rezende2015variational} or to investigate variable length input in distributed training as in \cite{ergen2017online}.

\bibliography{example_paper}
\bibliographystyle{icml2018}

\end{document}